\documentstyle[12pt]{article}

\newcommand{\beq}{\begin{equation}}
\newcommand{\eeq}{\end{equation}}
\newcommand{\bea}{\begin{eqnarray}}
\newcommand{\eea}{\end{eqnarray}}

\newcommand{\qs}{/\kern-.52em s}

\begin{document}
\title
{On the Electric Charge of Monopoles at Finite Temperature}
\author{
J.C. Le Guillou \thanks{Also at {\it Universit\'e de Savoie} and at
{\it Institut Universitaire de France}}\,
and
F.A. Schaposnik \thanks{Investigador CICBA, Argentina}
\thanks{On leave from Universidad Nacional de La Plata, Argentina}\\
Laboratoire de Physique Th\'eorique ENSLAPP 
\thanks{URA 1436 du CNRS associ\'ee
\`a l'Ecole Normale Sup\'erieure de Lyon et \`a 
l'Universit\'e de Savoie}\\
LAPP, B.P. 110, F-74941 Annecy-le-Vieux Cedex, France\\
}

\date{}

\maketitle
\thispagestyle{empty}

\def\thepage{\protect\raisebox{0ex}{\ } 
{{\small E}{{N}}{\large S}{\Large L}{\large A}{{P}}{\small P}}
-A-584/96}
\thispagestyle{headings}
\markright{\thepage}

\begin{abstract}
\normalsize
We calculate the electric charge at finite temperature $T$
for non-Abelian monopoles in
spontaneously broken gauge theories with a CP violating $\theta$-term. 
A careful treatment of 
dyon's gauge degrees of freedom shows that
Witten formula for the dyon charge at $T=0$, $ Q = e(n - \theta/2\pi) $,
remains valid at $T \ne 0$. 
\end{abstract}
\newpage
\pagenumbering{arabic}

Whenever CP invariance is violated by a $\theta$-term, 
dyons acquire an electric charge which
depends on
to the vacuum angle $\theta$ through the remarkable 
relation \cite{Wi}:
\beq
Q = -\frac{e \theta}{2 \pi} + n e
\label{uno}
\eeq
Here $e$ is the unit electric charge and $n$ an integer.
This relation was originally obtained, starting from an $SO(3)$
gauge
theory spontaneously broken to $U(1)$ by the
vacuum expectation value of a Higgs field in the adjoint
representation,
using semiclassical 
arguments and also by canonical methods. Following this last 
approach, one defines the operator $N$ that generates 
gauge transformations around the $U(1)$ 
(electromagnetic) surviving symmetry
and then imposes,
as an operator statement,
\beq
\exp(2\pi i N) = I
\label{dos}
\eeq
Now, when a CP violating term of the form 
\beq
\Delta{\cal L} = \theta \frac{e^2}{32 \pi^2} 
{^*}\!F_{\mu \nu}F^{\mu \nu}
\label{3} 
\eeq
is added
to the original Lagrangian, 
one can see that condition (\ref{dos}) implies relation (\ref{uno}).

Afterwards there have been several investigations
of fermion-monopole systems leading to the determination
of the monopole quantum numbers. According to  different
assumptions on boundary conditions for the Dirac equation,
different results were obtained for the fermion contribution
to the monopole electric charge \cite{Callan}-\cite{Gold}.

Gauge invariance and topological
considerations are
the basic ingredients for deriving formula (\ref{uno})
much in the same spirit as 
they are in the arguments leading 
to quantization of the Chern-Simons coefficient
both at the classical and quantum level
\cite{Jac}-\cite{Al}. It is then natural to
expect that eq.(\ref{uno}) does not get 
smoothly renormalized at finite temperature.
An analysis of an abelian fermion-Dirac monopole system
at finite temperature reported in these pages
\cite{CP} seems to indicate, however, that 
at finite temperature
the dyon's charge 
is no more quantized according to eq.(\ref{uno}). As the
authors already remarked, their analysis is connected 
with that leading to a temperature dependent coefficient
for the Chern-Simons effective action which arises
when considering three-dimensional fermions.
Now, it has been shown in refs.\cite{Pi}-\cite{BFS}
that the requirement of gauge invariance 
together with topological reasons impose
severe constraints on the temperature dependence
of the induced charge in this last system: 
it can be at most an integer
function of the temperature so that any smooth
behavior 
of the induced charge with the temperature
is excluded. One can then think that 
results of \cite{CP}
indicating a temperature dependent change
in formula (\ref{uno}) are the reflection
of the above mentionned
subtleties already encountered at $T = 0$
for fermion monopole systems
as well as those arising in the analysis of
Chern-Simons theories at finite temperature.

It is then an interesting open
question, already stressed
in \cite{CP}, whether formula (\ref{uno}), which is known to hold
for non-Abelian monopoles in
spontaneously broken gauge theories with a CP violating term 
(and no fermions), is modified at finite temperature.
It is the purpose of the present work to discuss this issue
showing that a careful treatment of gauge degrees of freedom
leads to
condition (\ref{uno}) both at $T = 0$ and at $T \ne 0$.

We start for definiteness with an $SO(3)$ gauge theory 
spontaneously
broken to $U(1)$ by the vaccum expectation value of a Higgs 
field ${\vec \phi}$ in the adjoint representation. 
The corresponding Lagrangian reads
\beq
{\cal L} = -\frac{1}{4}  {\vec F}_{\mu \nu}^2 + \frac{1}{2}  
({ {D_\mu {\vec\phi}}})^2
- V[\vert{\vec \phi}\vert]
\label{luno}
\eeq
here $V[\vert{\vec \phi}\vert]$ is a symmetry
breaking potential having its minima
at ${\vec \phi} = {\vec \phi}_0$ and
\beq
D_{\mu}{\vec \phi} = \partial_\mu {\vec \phi} +{e}
{\vec A_\mu} \wedge {\vec \phi}
\label{derco}
\eeq
As it is well known  
this Lagrangian admits monopole-like
solutions \cite{tH}-\cite{P} with quantized magnetic charge
and, at the classical level, arbitrary  electric  charged \cite{JZ}.
At the quantum level, it has been proven that
the dyon electric charge $Q$ is quantized to be an integer multiple
of the fundamental charge $e$, $Q = ne$ \cite{GJ}. As stated
above, the situation radically changes when a CP violating 
term of the
form (\ref{3}) is added to Lagrangian (\ref{luno})
\beq
L = {\cal L} + \frac{\theta e^2}{32\pi^2}{^*{\vec F}^{\mu \nu}}
. {{\vec F}_{\mu \nu}}
\label{ldos}
\eeq
with $^*F^{\mu \nu} = (1/2)\epsilon^{\mu \nu \alpha \beta} 
F_{\alpha \beta}$.
It is precisely the finite
temperature behavior of the system with dynamics 
governed by Lagrangian (\ref{ldos})
that we shall study in what follows.

As stressed in ref.\cite{Callan} the electric degree of freedom
of the dyon corresponds essentially to a rigid rotator collective
coordinate associated with gauge transformations. Quantization
of the dyon charge will then result from quantization of the
rigid rotator energy levels. To see this in more detail let us
recall that the static, spherically
symmetric 't Hooft-Polyakov monopole solution
can be written in the form
\beq
 {\vec A}_i^{mon}({\vec x}) = \frac{1}{e}(K(r)-1) {\vec \Omega} \wedge 
 \partial_i{\vec \Omega}
\label{a}
\eeq
\beq
{\vec \phi}^{mon}({\vec x}) = \frac{1}{e} \frac{H(r)}{r} {\vec \Omega}
\label{f}
\eeq
Here the isovector ${\vec \Omega}$ is defined as
\beq
{\vec \Omega} =\frac{{\vec x}}{r}
\label{o}
\eeq
and $H(r)$ and $K(r)$  satisfy at the origin
$H(0) = 0$, $K(0) = 1$ while at infinity one has
$H(r)/r\vert_\infty = \vert {\vec \phi}_0 \vert$, 
$K(r)\vert_\infty = 0$ 
ensuring
that the monopole has unit magnetic charge.
 
The monopole can be endowed with electric charge by including
an ${\vec A}_0$ component \cite{JZ},
\beq
{\vec A}_0({\vec x}) = \frac{1}{er} J(r) {\vec \Omega}
\label{d}
\eeq
with $J(0) = 0$ and $J(r) \to Mr + b$ for $r \to \infty$. ($M$, 
which has the dimension of a mass, sets the scale for 
$J$ while $b$ determines the electric charge of the dyon). Indeed,
the $U(1)$ electromagnetic field, as identified by 't Hooft \cite{tH},
\beq
{\cal F}_{\mu \nu} = {\vec \Omega}. {\vec F}_{\mu \nu} - 
\frac{1}{e \vert {\vec \phi}\vert ^2} {\vec \Omega}.D_\mu {\vec \phi} 
\wedge 
D_\nu {\vec \phi}
\label{el}
\eeq
shows the existence of a radial electric field
\beq
{\cal F}_{0r} = - \frac{d}{dr} \left[\frac{J(r)}{er}\right]
\label{E}
\eeq
and an electric charge
\beq
Q \equiv \int dS_i {\cal F}_{0i} = \frac{ 4\pi b}{e}
\label{bb}
\eeq
The coupled Euler-Lagrange equations relate $b$ to the other
parameters of the theory in such a way that the  electric
charge remains unquantized at the classical level.

There is an alternative way of obtaining electrically
charged monopoles from 't Hooft-Polyakov
solution, which
turns to be more convenient concerning quantization.
Indeed, working in the ${\vec A}_0 = 0$
gauge, one can make arise the dyon degree of freedom 
\cite{Callan} by considering time-dependent transformations
$U({\vec x},t)$, leaving the Higgs field invariant, of the form
\beq
U[\lambda] = \exp[i \lambda(r,t) \vec\Omega . {\vec t}]
\label{U}
\eeq
which act on the $A_i's$ so that
\beq
{ A}_i^{mon}  \to  U[\lambda] {A}_i^{mon} 
U^{-1}[\lambda] + \frac{i}{e} U[\lambda] \partial_i U^{-1}[\lambda]
\label{cambio}
\eeq
Here $ t^a$ ($a =1,2,3$) are the $SO(3)$
generators, $t^a = \sigma^a/2$, and we write
$A_i = {\vec A}_i . {\vec t}$. Because of the underlying spherical 
symmetry,
$\lambda$ depends only on the radial coordinate $r$.
Note that $U$ does not correspond to
a full gauge transformation since we
continue to work
at fixed gauge ${A}_0 = 0$. Now, after change (\ref{cambio}) 
there is an electric field
whose radial component takes the form
\beq
{\cal F}_{0r} = \frac{1}{e} \frac{d{\dot\lambda}}{dr}
\label{ele}
\eeq
with $\dot \lambda = (d/dt)\lambda$ satisfying the equation
\beq
\Delta {\dot \lambda} = \frac {2}{r^2} {\dot \lambda} K^2
\label{ss}
\eeq
In particular, for
\beq
\lambda(r,t) = \alpha(t) \frac{J}{Mr}
\label{rela}
\eeq
(eq.(\ref{ss}) then being satisfied)
one obtains  the same radial component for the electric field as that
arising from the static Julia-Zee dyon, eq.(\ref{E}), provided 
one choses $\alpha(t) = -M t$ . Now, 
Lagrangian (\ref{luno})
becomes, after transformation (\ref{cambio})
\beq
{\cal L} = \frac{1}{2e^2} \partial^i({\dot \lambda}
\partial_i{\dot \lambda})
+ M_{mon}
\label{div}
\eeq
where $M_{mon}$ is the  't Hooft-Polyakov monopole mass.
At the quantum level, when working in
the monopole sector,  the term  $M_{mon}$ should be substracted
in order to normalize the zero-point energy so that the
monopole energy is equal to zero \cite{Ruba}.

Concerning the CP violating Lagrangian (\ref{3}) it becomes
\beq
\Delta{\cal L} = - \frac{\theta}{8 \pi ^2}
\frac{1}{r^2}\frac{d}{dr}[{\dot \lambda}(K^2(r) - 1)]
\label{33} 
\eeq
Using eqs.(\ref{div})-(\ref{33}) and treating $\lambda$ as 
a collective coordinate one can reobtain 
Witten's result (\ref{uno}) at zero temperature.
We shall not
perform this analysis here but 
directly consider the case $T \ne 0$.

Finite temperature calculations are carried out as usual by
compactifying the (Euclidean) ``time'' variable into the range
$0 \le \tau \le \beta = 1/T$ (in our units, $\hbar = c = k = 1$).
Periodic boundary conditions (in ``time'') have to be used for 
gauge and Higgs field. In particular \cite{GPY}

\beq
{\vec A}_i({\vec x},\beta) = {\vec A}^{\Lambda}_i({\vec x},0)
\label{ojo}
\eeq
That is, periodicity modulo $\tau$-independent  transformations
$\exp[i\Lambda({\vec x})]$ with
$\Lambda(\infty) = 0$ is imposed. Here, we have defined
\beq
{\vec A}^{\Lambda}_i({\vec x},0) = 
\exp[i\Lambda({\vec x})]{\vec A}_i({\vec x},0) 
\exp[-i\Lambda({\vec x})]
+ \frac{i}{e} \exp[i\Lambda({\vec x})] \partial_i 
\exp[-i\Lambda({\vec x})]
\label{agreg}
\eeq

The correct partition function at finite temperature is then given by
\begin{equation}
{\cal Z} =
  \int_{\Lambda(\infty)=0} D\Lambda 
  \int_{{\vec A}_i({\vec x},\beta) = 
  {\vec A}^{\Lambda}_i({\vec x},0)}
   {D}{\vec A}_i \> D{\vec \phi}
                  \exp\left( -\int_0^\beta d\tau \int d^3x L \right)
,
\label{Zufa}
\end{equation}
with $L$ the Euclidean version of
Lagrangian (\ref{ldos}). The $\Lambda$ integration takes into
account
the imposition of Gauss law on physical
states within the path-integral framework. As
explained above, the $A_i$ integration 
covers all fields periodic up to a twist. Concerning
the $\phi$ integration,
is should be performed over periodic field configurations 
(see ref.\cite{GPY} for a thoroughful discussion
on the definition of the partition function as in eq.(\ref{Zufa})).

To see how a careful handling of the dyon degree of freedom implies
its charge quantization according
to eq.(\ref{uno}) also at finite temperature, we follow
the same collective coordinate treatment 
leading to charge quantization at $T=0$
\cite{Callan}. To this end we write the path-integral variables
in the form 

\begin{eqnarray}
A_i({\vec x},\tau) & = & A^\lambda_i({\vec x},\tau) + a_i({\vec x},\tau)
\label{colT}\\
\phi({\vec x},\tau) & = &  \phi^{mon}({\vec x}) + \varphi 
({\vec x},\tau)\nonumber
\end{eqnarray}
with $A^\lambda_i({\vec x},\tau) $ the $U$-transformed 't Hooft-Polyakov
monopole solution,
\beq
A^\lambda_i({\vec x},\tau) 
= U[\lambda] A_i^{mon}({\vec x}) U^{-1}[\lambda]
+\frac{i}{e} U[\lambda] \partial_i U^{-1}[\lambda] ~,
\label{susi}
\eeq
\beq
U[\lambda] = \exp[i \lambda(r,\tau) \vec \Omega . \vec t]
\label{sot}
\eeq
(Collective coordinates associated to non-gauge symmetries will not
be considered in what follows since they do not play any role in 
the charge quantization problem). 
One has explicitely,
\begin{eqnarray}
e {\vec A}^\lambda_i & = & (\partial_i \lambda) {\vec {\Omega}}
+ K \sin \lambda \partial_i{\vec {\Omega}} + (K \cos \lambda - 1)
{\vec {\Omega}} \wedge \partial_i {\vec {\Omega}}
\nonumber\\
-ie {\vec F}^\lambda_{0i} & = & (\partial_i {\dot \lambda})
{\vec \Omega}
+ {\dot \lambda} K \cos \lambda \partial_i {\vec {\Omega}}
- {\dot \lambda}K \sin \lambda {\vec \Omega} \wedge \partial_i 
{\vec \Omega}
\nonumber\\
e {\vec F}^\lambda_{ij} & = & (K^2 - 1) \partial_i {\vec \Omega} 
\wedge
\partial_j {\vec \Omega} + \sin \lambda 
(\partial_i K\partial_j{\vec \Omega} - \partial_j 
K\partial_i{\vec \Omega})
\nonumber\\
& & + \cos  \lambda (\partial_i K {\vec \Omega} \wedge \partial_j
{\vec \Omega}
-\partial_j K 
{\vec \Omega} \wedge \partial_i{\vec \Omega})
\label{s3}
\end{eqnarray}
with now ${\dot \lambda} = (d/d\tau)\lambda$.

In terms of the new variables
the partition function reads
\beq
Z = \int D\Lambda \int Da_i D\varphi D\lambda \Delta
\exp (- \int_0^\beta d\tau
\int d^3x L[A^{mon},\phi^{mon},a_i,\varphi,\lambda]) ~,
\label{sui}
\eeq
With $\Delta$ the Jacobian associated
with the change to the new integration variables.
Now, using eqs.(\ref{div})-(\ref{33}) one has  
\beq
Z = Z_d \times Z_f
\label{pro}
\eeq
where $Z_d$ corresponds to the path-integral over
the collective coordinate $\lambda$ accounting
for the dyon's degree of freedom and is obtained using eq.(\ref{s3})
as
\begin{eqnarray}
Z_d & = & \int D\lambda \exp
\left(-\frac{2\pi}{e^2}\int_0^\beta d\tau \int_0^\infty r^2 dr 
\partial^i({\dot \lambda}\partial_i{\dot \lambda}) \right.
\nonumber\\
& & \left. -i \frac{\theta}{2 \pi}\int_0^\beta d\tau \int_0^\infty
dr \frac{d}{dr}[(K^2-1){\dot \lambda}] \right)
\label{fi}
\end{eqnarray}
(We have not explicitely written the $\Lambda$ integration which,
as we shall see below, will be taken
into account by giving a precise meaning to the
$\lambda$ integration). 
Concerning $Z_f$, it includes the contributions coming from
the classical action and the
integration over fluctuations $a_i$ and $\varphi$
not relevant for the analysis of the dyon charge.

In the $T = 0$ case \cite{Callan} one can identify 
the Minkowskian version of $Z_d$,  $Z^M_d$,
with 
the partition function for a one dimensional Coulomb system
with a position-dependent coupling (proportional
to $r^2$). The quantum analysis of such a system leads
to the usual spectrum of dyon states \cite{JZ}-\cite{GJ}
with charge given by formula (\ref{uno}) \cite{Wi}.
Alternatively, one can relate $Z^M_d$ with the partition
function for a quantum rotator \cite{Wil} and and also infer 
dyon charge quantization
from the energy spectrum of the rotator \cite{TK}. 

To extend this analysis to the case of finite temperature,
let us start by noting that, as in the $T=0$ case
\cite{Callan},
it will be sufficient to consider the 
collective coordinate $\lambda$ in the form
\beq
\lambda =\frac{J(r)}{Mr} \alpha(\tau)
\label{alT}
\eeq
with $J(r)$ giving  the (classical) electric potential of a Julia-Zee
dyon (see eqs.(\ref{E}),(\ref{rela})) and $\alpha(\tau)$
describing quantum effects at finite temperature. 

In order to unravel
the angular character of $\alpha$
let us come back to condition (\ref{ojo}) 
that $A^\lambda_i$ should also satisfy.
Since we have just considered a spherically symmetric
collective coordinate $\lambda$ in the form (\ref{alT}),
the allowed transformations 
for $A^\lambda_i$ should take the form $\Lambda({\vec x})
= \Lambda(r){\vec \Omega}.{\vec t}$ with
$\Lambda(r=\infty) = 0$. Explicitely one gets
\begin{eqnarray}
{({\vec A}^\lambda)}^\Lambda_i({\vec x},0) & = & 
\partial_i\left(\lambda(r,0) + \Lambda(r)\right)
+ K \sin (\lambda(r,0)+ \Lambda(r)) \partial_i{\vec \Omega} +
\nonumber\\
& &\left(K\cos (\lambda(r,0) + \Lambda(r)) - 1\right)
{\vec \Omega} \wedge \partial_i{\vec \Omega}
\label{uras1}
\end{eqnarray}
\begin{equation}
{\vec A}^\lambda_i({\vec x},\beta)  =  
\partial_i\lambda(r,\beta)
+ K \sin \lambda(r,\beta)  \partial_i{\vec \Omega} +
\left(K\cos \lambda(r,\beta) - 1\right)
{\vec \Omega} \wedge \partial_i{\vec \Omega}
\label{uras2}
\end{equation}
The ``periodicity'' condition for $A^\lambda_i$
thus leads to
\beq
\left(\alpha(0) - \alpha(\beta)\right)\frac{J(r)}{Mr} + \Lambda(r)
 = 2\pi n, ~~~ n \in {\cal Z}
\label{usas}
\eeq
Now, consistency of this condition at  at $r \to \infty$ 
implies
\beq
\alpha(0) = \alpha(\beta) + 2 \pi n
\label{texto}
\eeq
\beq
\Lambda_n(r) = 2\pi n (1 - \frac{J(r)}{Mr})
\label{cuci}
\eeq

We can then rewrite partition function $Z_d$ in terms of the
angular variable $\alpha(\tau)$ in the form
\beq
Z_d = \int D\alpha \exp[-\int_0^\beta
d\tau (\frac{{\cal J}}{2}{\dot \alpha}^2
- \frac{i\theta }{2\pi} {\dot \alpha})] 
\label{cerca}
\eeq
with ${\cal J} = -4\pi b/{e^2 M}$. In eq.(\ref{cerca}) it is 
understood
that the boundary condition (\ref{texto}) is taken into account
so that in fact the $\alpha$-integration covers the different
``$n$-sectors''
implied by the $\Lambda$ integration  included in the definition of
the finite temperature partition function (\ref{Zufa}).
Note that one can easily relate ${\cal J}$ with the dyon electric 
charge
by using eqs.(\ref{colT})-(\ref{alT}):
\beq
\frac{Q_{el}}{e} =  
\frac{1}{e \vert {\vec \phi}_0 \vert } 
\int d^3x \partial^i({\vec \phi}.{\vec F}^{\lambda}_{0i}) =
i {\cal J} {\dot \alpha }
\label{carga}
\eeq
We recognize in eq.(\ref{cerca}) the path-integral expression
for the transition amplitude of a (modified) rigid rotator
evolving (in Euclidean ``time'' $\tau$)
from an ``initial'' state $\alpha' = \alpha(0)$
to a ``final'' state $\alpha'' = \alpha(\beta)$ modulo $2 \pi$
which should be identified with $\alpha'$
according to condition (\ref{texto}).
Now, the spectrum of such a
rotator can be derived without
leaving the  path-integral framework as follows
\cite{Zinn}. Consider that the ``time'' interval is $[0,\beta]$
and first solve the classical
equations of motion for $\alpha$ as
\beq 
\alpha_{sol}^{(n)}(\tau) = \alpha' +  \frac{\tau}{\beta}(\alpha'' - 
\alpha' +2\pi n),
\label{sol}
\eeq
One then writes the path-integral variable in each ``$n$-sector''
in the form
\beq
\alpha(\tau) = \alpha_{sol}^{(n)}(\tau) + u(\tau)
\label{ka}
\eeq
so that $Z_d$ can be written as
\beq
Z_d = Z_d[\alpha',\alpha']
\label{traza}
\eeq
with
\beq
Z_d[\alpha',\alpha''] = {\cal N}(\beta)\sum_{n = - \infty}^{\infty} 
\exp[-\frac{\cal J}{2 \beta}(\alpha'' - \alpha' +2\pi n)^2 + 
i\frac{\theta}{2\pi}(\alpha'' - \alpha' +2\pi n)]
\label{roti}
\eeq
Here ${\cal N}(\beta)$ gives a  normalization
factor arising from integration over fluctuations around the classical
trajectories, ${\cal N}(\beta) = \sqrt{2\pi{\cal J}/ \beta}$. 
One now proceeds to a Fourier expansion
\beq
Z_d[\alpha',\alpha''] = \sum_{l = -\infty}^{\infty} 
\exp[ i l (\alpha'' - \alpha') ]
\exp(-\beta E_l)
\label{Four}
\eeq
with
\begin{eqnarray}
\exp{(-\beta E_l)}& & = 
\sqrt{\frac{\cal J}{2\pi \beta}}\int_0^{2\pi} d\alpha
\exp(-i l \alpha) \times\nonumber\\
& & \sum_{n = -\infty}^{\infty}\exp[-\frac{\cal J}{2\beta}(\alpha 
 +2\pi n)^2 + 
i\frac{\theta}{2\pi}(\alpha + 2\pi n)]
\label{pa}
\end{eqnarray}
Inverting sum and integration and making the shift 
$\alpha + 2\pi n \to \alpha$ one finally gets
\beq
\exp[-\beta E_l] =  
\exp[-\frac{\beta(l-{\theta }/{2\pi})^2}{2 {\cal J}}]
\label{ufa}
\eeq
thus giving for the energy levels of the rotator the result
\beq
E_l = \frac{1}{2{\cal J}}(l - \theta /2 \pi)^2
\label{ene}
\eeq
We can go further in the analogy with a quantum rigid rotator 
and evaluate
from the Euclidean Lagrangian in partition function (\ref{cerca})
the (angular) momentum $p_\alpha$
\beq
p_\alpha \equiv i\frac{\partial L}{\partial{\dot \alpha}} = 
i{\cal J}{\dot \alpha} + \frac{\theta}{2\pi} =
\frac{Q_{el}}{e} + \frac{\theta}{2\pi} 
\label{lio}
\eeq
On the other hand, eq.(\ref{ene}) shows that $p_\alpha$ 
is quantized according to
$p_\alpha = l$ . From this
one finally has the dyon charge quantization condition
\beq
Q_{el} = e(l - \theta/2\pi), ~~~ l \in {\cal Z}
\label{unoo}
\eeq
One arrives in this way to Witten formula (\ref{uno})
for the dyon charge, which thus
remains unchanged at finite temperature. To corroborate
this result obtained just by using the quantum-mechanics analogy,
one can directly evaluate
\beq
<i{\cal J}{\dot \alpha} + \frac{\theta}{2\pi}> = \frac{1}{Z_d}
\int  D\alpha (i{\cal J}{\dot \alpha} + \frac{\theta}{2\pi})
\exp[-\int_0^\beta d\tau (\frac{{\cal J}}{2}{\dot \alpha}^2
-i \frac{\theta }{2\pi} {\dot \alpha})]  
\label{cercas}
\eeq
Proceeding as before, we find 
\begin{eqnarray}
<i{\cal J}{\dot \alpha} + \frac{\theta}{2\pi}> & = & 
<\frac{Q_{el}}{e} + \frac{\theta}{2\pi}>\nonumber\\
& = & \frac{1}{Z_d}\sum_l \,l \,
\exp(-\beta E_l)
\label{sera}
\end{eqnarray}
which again leads to quantization of $p_\alpha$ 
as an integer $l$ and thus
to eq.(\ref{unoo}).

~

In summary, our analysis indicates that the formula
(\ref{uno}) found in \cite{Wi} for the electric charge of a 
't Hooft-Polyakov magnetic monopole in CP non-conserving
spontaneously broken gauge theories remains unchanged 
at finite temperature.
This result should be compared with that obtained in \cite{CP}
for a fixed Dirac monopole in an electron-positron plasma
at temperature $T$. In this last investigation
the induced charge does depend smoothly on the temperature
and on the fermion  mass $m$  through the dimensionless
combination $m/T$ (being their monopole an Abelian
Dirac one, there is no other mass parameter
to enter in a dimensionless variable describing
the temperature dependence). This behavior  can be related,
as noted in \cite{CP} to perturbative
analysis leading
to a temperature dependent coefficient for the
Chern-Simons effective action in $3$-dimensional 
fermionic theories. However, it was shown in
\cite{Pi}-\cite{BFS} that perturbative approaches yielding
to this last result are inconsistent
with the requirement of gauge-invariance. 
Moreover, already at $T=0$ the fermion-Dirac monopole system
has shown to lead to different monopole quantum numbers
according to different boundary condition
for the Dirac wave function at the location of the
monopole. As it is the
case for Witten analysis \cite{Wi}, our treatment
stems from more general grounds and applies to extended
't Hooft- Polyakov monopoles in
spontaneously broken gauge theories
where temperature dependence could 
in principle arise in terms of a dimensionless
variable, for example the
combination $m_W/T$ (with $m_W$ the mass of the 
$W$ vector boson). However, 
because of gauge-invariance and topological considerations,
already at the root of eq.(\ref{uno}) at $T=0$, 
we have shown that
the dyon electric charge remains unchanged at finite temperature.

~~

\underline{Acknowledgements}:
This work was
supported in part by  CICBA and CONICET, Argentina.
F.A.S. ~ thanks the La\-bo\-ra\-toire de Phy\-si\-que 
Th\'eo\-ri\-que
ENSLAPP for its kind hospitality.

\end{document}